# Quantifying 'word salad': The structural randomness of verbal reports predicts negative symptoms and Schizophrenia diagnosis 6 months later

## - Thought disorder measured as random speech structure -

Natália B. Mota (1), Mauro Copelli (2), Sidarta Ribeiro (1)

((1) Brain Institute, Federal University of Rio Grande do Norte – UFRN, (2) Department of Physics, Federal University of Pernambuco – UFPE)

**Background:** The precise quantification of negative symptoms is necessary to improve differential diagnosis and prognosis prediction in Schizophrenia. In chronic psychotic patients, the representation of verbal reports as word graphs provides automated sorting of schizophrenia, bipolar disorder and control groups based on the degree of speech connectedness. Here we aim to use machine learning to verify whether speech connectedness during first clinical contact can predict negative symptoms and Schizophrenia diagnosis six months later.

**Methods:** PANSS scores and memory reports were collected from 21 patients undergoing first clinical contact for recent-onset psychosis and followed for 6 months to establish DSM-IV diagnosis, and 21 healthy controls. Each report was represented as a graph in which words corresponded to nodes, and node temporal succession corresponded to edges. Three connectedness attributes were extracted from each graph, z-scores to random graph distributions were measured, correlated with the PANSS negative subscale, combined into a single Fragmentation Index, and used for predictions.

**Findings:** Random-like speech was prevalent among Schizophrenia patients (64% x 5% in Control group, p=0.0002). Connectedness explained 92% of the PANSS negative subscale variance (p=0.0001). The Fragmentation Index classified low versus high scores of PANSS negative subscale with 93% accuracy (AUC=1), predicted Schizophrenia diagnosis with 89% accuracy (AUC=0.89), and was validated in an independent cohort of chronic psychotic patients.

**Interpretation:** The structural randomness of speech graph connectedness is increased in Schizophrenia. It provides a quantitative measurement of "word salad" as a Fragmentation Index that tightly correlates with negative symptoms and predicts Schizophrenia diagnosis during first clinical contact of recent-onset psychosis.

Address correspondence to Natália Bezerra Mota,

Brain Institute - Federal University of Rio Grande do Norte (UFRN): Av Nascimento de Castro 2155, CEP: 59056-450, Natal, RN, Brazil; Phone: +55 84 32152708

e-mail: nataliamota@neuro.ufrn.br;

**Background**

Schizophrenia is associated with negative symptoms and poor prognosis [1]. In particular, elevated negative symptoms are associated with lower rates of recovery [1,2]. Formal thought disorder - which comprises poverty of speech, derailment, and incoherence - constitutes an important set of psychotic symptoms, and negative formal thought disorder is associated with the Schizophrenia diagnosis, even during first episode psychosis [2,3]. The early stages of the disease constitute a critical opportunity for prevention of major cognitive damage [4].

Improved behavioral measures subjected to novel mathematical analyses are emerging as part of a new field that uses computational tools to better characterize psychiatric phenomena (computational phenotyping) [5-13]. In particular, verbal report quantification by graph tools predicts diagnosis [9,10] and makes precise and automated quantification of speech features related with negative symptoms [9]. By representing each word as a node and the temporal sequence of consecutive words as directed edges, it is possible to calculate attributes that characterize graph structure. The assessment of dream reports from chronic psychotic patients has shown that word connectedness (number of edges between words, or amount of nodes in connected components) is negatively correlated with negative symptoms [9]. The same graph attributes calculated from short-term memory reports from healthy children are positively correlated with Intelligence Quotient and Theory of Mind scores, and can predict academic performance independently of other cognitive measures [14]. Altogether, the data suggest that word connectedness measured by graph analysis rises during healthy development, but not during the course of Schizophrenia. It is therefore possible that early markers of speech disorganization during recent-onset psychosis, such as decreased connectedness, may be able to predict the severity of negative symptoms as well as the Schizophrenia diagnosis.

Our hypotheses were: 1. Speech connectedness from dream reports [9] and short-term memory reports [14] can predict the Schizophrenia diagnosis ; 2. Patients in the Schizophrenia group will produce verbal reports less connected and more similar to random connectedness then those from other groups; 3. Connectedness attributes negatively correlate with negative symptoms measured by PANSS [9]; 4. A single index combining connectedness attributes correlated with negative symptoms should improve Schizophrenia diagnosis and prediction of negative symptom severity.

## Methods

**Study Design:**

This prospective study recruited patients interviewed during first clinical contact for recent-onset psychosis in a public child psychiatric clinic (CAPSi) in Natal during 2014. Patients were followed up for 6 months to establish diagnoses according to DSM IV criteria [15]. After this sample was collected, well-matched controls were recruited on nearby public schools. Data analysis began after the entire sample was collected and all patients had finished follow up.

**Participants:**

Study approved by the UFRN Research Ethics Committee (permit # 742-116). Twenty-two patients undergoing recent-onset psychosis (Table 1) were recruited during first psychiatric interview and followed for 6 months to establish diagnoses by the entire clinical team. Inclusion criterion was to be seeking treatment for psychotic symptoms for the first time (maximum duration of two years as reported by patient and family members). Exclusion criteria comprised having any neurological symptom, or having drug-related disorders. One patient was excluded after epilepsy diagnosis. Twenty-one healthy control subjects matched by age, sex and education were interviewed during regular class time in public schools of Natal (Table 1). An additional exclusion criterion for the Control group was not having any psychiatric symptom or diagnosis, as assessed during family member interviews. Participants and legal guardians provided written informed consent.

**Protocol:**

Subjects were submitted to an audio-recorded interview that consisted of requests for 6 time-limited memory reports. In order to minimize inter-subject differences in word count, each report was limited to 30 seconds. Whenever the subject spontaneously stopped the report, he/she was stimulated to keep talking by way of general instructions like "please, tell me more about it". When the report reached the 30 seconds limit, the interviewer interrupted the report saying "ok". The interview began with a request to produce a "*dream report*" (either recent or remote). Next, the "*oldest memory report*" was obtained by requesting the subjects to report the most remote memory they could access at that moment. Then the subjects were requested to report on their previous day ("*yesterday report*"), and finally they were exposed to 3 images presented on a computer screen, comprising a "*highly negative image*", a "*highly positive image*" and a "*neutral image*" from the IAPS database [16] of affective images, previously tested in children [16] and psychotic subjects [17]. Subjects were instructed to pay attention to each image for 15 seconds and then report an imaginary story based on it. The entire memory report protocol took up to 10 minutes to be completed. Subjects undergoing recent-onset psychosis were then evaluated psychiatrically using the psychometric scale PANSS [18] composed of three subscales (positive, negative and general). The negative subscale measured 7 symptoms: Blunted affect (N1), Emotional withdrawal (N2), Poor rapport (N3), Passive/apathetic social withdrawal (N4), Difficulty in abstract thinking (N5), Lack of spontaneity and flow of conversation (N6), Stereotyped thinking (N7) [18].

**Graph measures:**

The search for a predictive index of connectedness was exploratory, and for that we tested 6 different kinds of memory reports, searching for the best combination of connectedness attributes. Memory reports were transcribed and represented as graphs in which each word was represented as a node, and the temporal sequence between consecutive words was represented by directed edges (Figure 1A) using the software *SpeechGraphs* (http://www.neuro.ufrn.br/softwares/speechgraphs) [9]. Three connectedness attributes were calculated: Edges (E), which measures the amount of links between words; largest connected component (LCC), which measures the amount of nodes in the largest component in which each pair of nodes has a path between them; and largest strongly connected component (LSC), which counts the amount of nodes in the largest component in which each pair of nodes has a mutually reachable path, i.e., node "a" reaches node "b" and node "b" reaches node "a" (Figure 1A).

**Analyses:**

Since formal thought disorder symptoms are frequently described as 'word salad', i.e. as random-like speech, we compared each memory report graph to 1,000 random graphs built with the same nodes and number of edges, but with a random shuffling of the edges that amounts to shuffling word (Figure 1B). Next we estimated the LCC and LSC z-scores between each original graph and the corresponding random graph distribution (Figure 1C). These normalized attributes were termed LCCz and LSCz. Formally, $LCCz = (LCC - LCCmr) / LCCsdr$ and $LSCz = (LSC - LSCmr) / LSCsdr$, with LCCmr and LSCmr corresponding respectively to mean LCC and LSC values in the random graph distributions; likewise, LCCsdr and LSCsdr denote the standard deviation of LCC and LSC from the random graph distribution. A graph was considered random-like when its connectedness attributes fell within 2 standard deviations from the mean of the random distribution (Figure 1D).

To avoid over-fitting and better combine the most informative connectedness attributes, we first applied 5 connectedness attributes (E, LCC, LSC, LCCz and LSCz) from each memory report as inputs to a Naïve Bayes classifier with cross-validation (10-fold) implemented with Weka software [19], and trained for the binary choice between the Schizophrenia group versus the sum of Bipolar and Control groups, using as golden standard the diagnostic reached after 6 months of follow-up. Classification quality was assessed using Accuracy (Acc, percentage of correctly classified subjects) and area under the receiver operating characteristic curve (AUC). A threshold of Acc = 75% correct or AUC = 0.75 was established in order to consider a memory report informative (Figure 2A, Table 2). Using Pearson correlations, we related each connectedness attribute from each informative memory report to the PANSS negative subscale (Figure 2B, Table 3), and compared the groups applying Kruskal-Wallis and Wilcoxon Ranksum test (Figure 2C, Table 4). All statistical analyses were corrected for multiple comparisons (Bonferroni). We selected only those connectedness attributes that presented any significant statistical difference between groups, and were also correlated with negative symptoms following Bonferroni correction. After selecting the most informative connectedness attributes, we combined and correlated them with the total score of the PANSS negative subscale using multilinear regression (Figure 3A). Attribute coefficients were extracted and this linear combination was used to create an index called "Fragmentation Index" (equation described in Findings Session). We also verified whether the Fragmentation Index differed between the groups using Kruskal-Wallis and Wilcoxon Ranksum tests (Figure 3B, Table 5). All statistical analyses used Matlab Software.

To verify whether the Fragmentation Index could predict the Schizophrenia diagnosis using only connectedness attributes from memory reports recorded during the first psychiatric interview, we used the binary classifier Naïve Bayes [19], and applied cross-validation (10-fold) to sort the patients that 6 months later received the Schizophrenia diagnosis from other groups. To verify whether the Fragmentation Index could correctly sort patients with severe negative symptoms from those with milder negative symptomatology, we divided the sample in 2 subsamples with high (more than the median) and low (less or equal the median) scores of total PANSS negative subscale. Next we verified whether the Naïve Bayes classifier was able to classify both samples using only the Fragmentation index. Classification quality was verified by measuring true positive rate (sensitivity), true negative rate (1-specificity), precision, recall, f-measure, AUC and Acc (Figure 3C, Table 6). Finally, we validated the Fragmentation Index obtained in the recent-onset psychosis sample (i.e. with the same coefficients) to a previously collected sample of dream reports from chronic psychotic subjects and matched controls (subjects diagnosed with Schizophrenia n=20, Bipolar Disorder n=20 and 20 subjects without psychosis according DSM-IV) [9]. As this previous protocol was not time-limited, verbosity differences were controlled using average graph attributes from 30 word-graphs (see [9] for details).

**Findings**

After 6 months follow-up, 11 of the 21 patients were diagnosed with Schizophrenia disorder, and 10 with Bipolar disorder (Table 1, Figure 4). These groups did not show any significant difference regarding demographical characteristics (age, sex, educational level, and family income) or disease duration (Table 1, Figure 4). The Schizophrenia group used more atypical antipsychotic medications and less mood stabilizers than the Bipolar group (Table 1).

Only dream reports and negative image reports allowed to predict the Schizophrenia diagnosis 6 months in advance against other conditions (Bipolar disorder or Control), with AUC > 0.75 and Acc > 75% correct using all the connectedness attributes (Figure 2A, Table 2). Dream reports yielded better prediction (AUC = 0.84, Acc = 81% correct) than negative image reports (AUC = 0.78, Acc = 76% correct; Figure 2A, Table 2), but some subjects were unable to recall a dream during their first interview (36% of the Schizophrenia group (N = 4), 20% of the Bipolar group (N = 2) and none of the Control subjects failed to recall a dream; Figure 4). For this reason, further analyses used only these 2 report types.

As predicted, Schizophrenia subjects produced less connected reports than subjects from other groups (Figure 2C, Table 4), but only negative image reports showed high similarity with random connectedness (LSCz were smaller for Schizophrenia group compared to Control group, Wilcoxon Ranksum test p = 0.0033, Figure 2C). Negative image reports from Schizophrenia subjects showed random-like connectedness more frequently than reports from the Control group (64% of Schizophrenia group versus 5% of Control group, Chi-square test p = 0.0002; Figure 1D). Reports from Bipolar subjects showed intermediate random-like connectedness (30%; Figure 1D).

In further agreement with our prediction, connectedness attributes were negatively correlated with the PANSS negative subscale (Figure 2B, Table 3). Connectedness attributes from negative image reports were more correlated with negative symptoms than connectedness attributes from dream reports (Figure 2, Table 3). For dream reports, only the PANSS N2 symptom (emotional withdrawal) was significantly - and negatively - correlated with E and LCC (Figure 2, Table 3). For negative image reports, total as well as negative subscale PANSS symptoms (with the exception of N5) were significantly and negatively correlated with nearly all the connectedness attributes, except E (Figure 2, Table 3).

Next we combined all the connectedness attributes that showed both significant differences from other groups, and significant correlations with negative symptoms. Multilinear correlations were calculated between total PANSS negative subscale and 5 attributes from both memory reports (LCC, LSC and LSCz from negative image reports; E and LCC from dream reports), or 3 attributes exclusively from negative image or dream reports. The combination of connectedness attributes from both kinds of reports explained nearly all the variance in total negative symptoms ($R^2$ = 0.92, p = 0.0001, Figure 3A), while using only negative image reports explained substantially less ($R^2$ = 0.74, p < 0.0001, Figure 3A), and using only dream reports even less ($R^2$ = 0.50, p < 0.0155, Figure 3A). The following equations defined "Fragmentation Indices" for either a combination of dream and negative image reports, or separately for negative image or dream reports:

Fragmentation Index (Negative + Dream) = 30.53 + LCC_negative * (−0.11) + LSC_negative *(0.15) + LSCz_negative*(−2.4) + E_dream * (−0.23) + LCC_dream * 0.16

Fragmentation Index (Negative) = 31.43 + LCC * (−0.3) + LSC * (0.08) + LSCz * (−2.1)

Fragmentation Index (Dream) = 27.15 + E * (-0.156) + LCC * (-0.08)

The Schizophrenia group showed higher Fragmentation Index than other groups using both reports (Kruskal-Wallis p = 0.0022, Figure 3B, Table 5), using only negative image reports (Kruskal-Wallis p = 0.0044, Figure 3B, Table 5), or only dream reports (Kruskal-Wallis p = 0.0140, Figure 3B, Table 5). The Fragmentation Index from both kinds of reports predicted the Schizophrenia diagnosis with AUC = 0.89 and Acc = 89%, and predicted the negative symptoms severity with AUC = 1 and Acc = 93% (Figure 3C, Table 6, Figure 4). The Fragmentation Index calculated exclusively from negative image reports predicted the Schizophrenia diagnosis with AUC = 0.78 and Acc = 79%, and predicted negative symptom severity with AUC = 0.97, Acc = 90% (Figure 3C, Table 6, Figure 4). The Fragmentation Index using only dream reports was also quite predictive (Schizophrenia diagnosis AUC = 0.81, Acc = 78%; Negative symptom severity AUC = 0.78, Acc = 80%; Figure 3C, Table 6, Figure 5).

To validate the method in an independent cohort, the Fragmentation Index was applied to dream reports of a previously collected chronic psychotic sample. The statistical differences among the groups resembled those found in the recent-onset psychosis sample (Kruskal-Wallis p < 0.0001, Figure 3D, Table 5), and the Fragmentation Index successfully predicted the Schizophrenia diagnosis (AUC = 0.81, Acc = 82%) and negative symptom severity (AUC = 0.84, Acc = 75%; Figure 3C, Table 6, Figure 5).

Importantly, there were no statistically significant differences between the Bipolar and the Control groups for any connectedness attribute, either in isolation or combined into a Fragmentation Index (Tables 4 and 5). However, in the Bipolar group of the chronic sample, the Fragmentation Index was higher than in the Control group (Wilcoxon ranksum p = 0.0060), although not sufficient for good sorting (AUC = 0.70, Acc = 70%).

## Interpretation

One of the promises of computational psychiatry is to provide quantitative phenotyping of relevant psychiatric symptoms [5-7,20]. Here we showed that speech graph analysis allows for the structural quantification of formal thought disorder, mathematically defined by the linear combination of connectedness graph attributes and their degree of similarity to randomly generated graph attributes. This procedure offers unbiased and precise numbers to what was previously only described by words. All the hypotheses raised were verified. First, dream and negative image reports were optimal to predict Schizophrenia diagnosis 6 months in advance. Connectedness attributes from dream reports were most predictive of Schizophrenia, with better performance than connectedness attributes from waking reports [9]. However, the difficulty shown by some subjects to recall dreams is a practical clinical concern that we sought to circumvent using short-term memory reports based on affective images [14]. As predicted, short-term memory reports were more informative than long-term memory reports ("yesterday" or "oldest" memories).

Estimation of randomness degree provides a quantitative measurement of 'word salad' at the structural level. Connectedness is often impaired in Schizophrenia subjects, to the point of being undistinguishable from random values. The results also confirmed that connectedness is negatively correlated with negative symptom severity. A combination of connectedness attributes explained nearly all the variance of the negative symptoms severity, reaching high classification accuracy for negative symptom severity and prediction of Schizophrenia diagnosis 6 months in advance. Graph analysis is a fast and low-cost tool for complementary psychiatric evaluation: Recording of 2 time limited memory reports takes ~3 min, audio transcription takes ~10 min, and data processing from text transcript to graph analysis is nearly instantaneous [9]. Whenever a patient fails to recall a dream, it is still possible to calculate an accurate Fragmentation Index using only a negative image report.

Of note, Bipolar and Control groups could not be differentiated in the recent-onset psychosis sample using neither connectedness attributes nor Fragmentation Index, but a significant difference was observed in the chronic sample. This may reflect the progressive cognitive deterioration in Bipolar patients, in contrast with the early onset of such symptoms in Schizophrenia [21]. Semantic computational strategies are likely to better to predict psychotic breaks during prodromal stages [11], or to differentiate patients with Bipolar Disorder from healthy controls [22].

Our study has limitations. First, a larger N is necessary to obtain more representative psychopathological boundaries for the Fragmentation Index, i.e. more reliable estimations of the linear combination coefficients. Second, the findings must be replicated with native speakers of other languages. Third, the medications taken by the Schizophrenia and Bipolar groups could not be matched due to treatment differences between pathologies, and the non-interventional experimental design. Fourth, the duration of psychotic symptoms before the first clinical interview was estimated by interviews with family and patient, and therefore was not precisely measured [23]. Fifth, a longitudinal prodromic evaluation is in order to describe how graph attributes progress over time in relation to clinical evolution, and how sensitive these attributes are to medication changes.

**Funding**

Work supported by UFRN, Conselho Nacional de Desenvolvimento Científico e Tecnológico (CNPq), grants Universal 480053/2013-8 and Research Productivity 306604/2012-4 and 310712/2014-9; Coordenação de Aperfeiçoamento de Pessoal de Nível Superior (CAPES) Project ACERTA; Fundação de Amparo à Ciência e Tecnologia do Estado de Pernambuco (FACEPE); FAPESP Center for Neuromathematics (grant # 2013/07699-0, São Paulo Research Foundation).

**Authors' contribution**

NBM performed data collection and figure preparation. NBM, MC and SR contributed study design, literature search, data analysis, data interpretation, and writing.

**Acknowledgements**

We thank CAPS Infantil Natal/RN for access to the patients; Diego Slezák, Cláudio Queiroz, Sandro de Souza and Mariano Sigman for insightful discussions; Débora Koshiyama for bibliographic support; Pedro PC. Maia, Gabriel M. da Silva and Jaime Cirne for IT support. In memory of Raimundo Furtado.

## Tables:

**Table 1:** Socio-economic and clinical information of Schizophrenia, Bipolar and Control groups. Mean ± standard deviation of age in years, family income in USD per month, educational level in years, disease duration in days. Percentage of male and female subjects per group and of subjects under specific types of medication. P values of Wilcoxon-Ranksum test or Chi-square test between Schizophrenia versus Bipolar and Control groups (general information) or Schizophrenia versus Bipolar group (clinical information) (group label according diagnosis from 6 months follow up).

| Demographic Characteristics | | Schizophrenia | Bipolar Disorder | Control | P Value S x (B+C) |
|---|---|---|---|---|---|
| Age (years) | | 14.64 ± 2.57 | 15.30 ± 3.77 | 15.43 ± 3.55 | 0.5837 |
| Family Income (U$ per month) | | 1304.55 ± 762.31 | 1190.00 ± 667.76 | 368.42 ± 151.76 | 0.3746 |
| Sex | Male | 82% | 27% | 45% | 0.0542 |
| | Female | 18% | 73% | 55% | |
| Years of Education (years) | | 5.73 ± 2.34 | 6.40 ± 3.77 | 8.05 ± 2.77 | 0.0810 |
| Psychiatric Assessment | | Schizophrenia | Bipolar Disorder | | P Value: S x B |
| Medication | Typical Antipsychotic | 55% | 60% | | 0.8008 |
| | Atypical Antipsychotic | 82% | 40% | | 0.0487 |
| | Mood Stabilizer | 9% | 70% | | 0.0041 |
| | Benzodiazepine | 9% | 10% | | 0.9435 |
| | Antidepressants | 9% | 20% | | 0.4755 |
| Disease Duration (days) | | 339.36 ± 244.80 | 370.60 ± 306.08 | | 1 |

**Table 2:** Classification quality to classify Schizophrenia group from others subjects using all 5 connectedness attributes (E, LCC, LSC, LCCz, LSCz) with different time-limited memory reports.

| Groups | Sensitivity | Specificity | Precision | Recall | F-Measure | AUC | Accuracy |
|---|---|---|---|---|---|---|---|
| Dream | 0.81 | 0.85 | 0.87 | 0.81 | 0.82 | 0.84 | 80.56 |
| Negative | 0.76 | 0.68 | 0.78 | 0.76 | 0.77 | 0.78 | 76.19 |
| Positive | 0.69 | 0.77 | 0.79 | 0.69 | 0.71 | 0.74 | 69.05 |
| Neutral | 0.69 | 0.71 | 0.76 | 0.69 | 0.71 | 0.63 | 69.05 |
| Yesterday | 0.69 | 0.54 | 0.70 | 0.69 | 0.70 | 0.64 | 69.05 |
| Oldest | 0.57 | 0.56 | 0.66 | 0.57 | 0.60 | 0.62 | 57.14 |

**Table 3:** Pearson correlation between connectedness attributes (E, LCC, LSC, LCCz, LSCz) and negative symptoms measured by PANSS (total negative subscale, N1, N2, N3, N4, N5, N6, N7), using dreams or negative image reports. Showed R, $R^2$ and p values (in bold significant results after Bonferroni correction for 80 comparisons – 5 attributes * 2 reports * 8 symptoms, $p < 0.0006$)

| Dream Reports | E | | | LCC | | | LSC | | | LCCz | | | LSCz | | |
|---|---|---|---|---|---|---|---|---|---|---|---|---|---|---|---|
| PANSS Negative Subscale | R | $R^2$ | p | R | $R^2$ | p | R | $R^2$ | p | R | $R^2$ | p | R | $R^2$ | p |
| Total | -0.71 | 0.50 | 0.0032 | -0.70 | 0.49 | 0.0038 | -0.63 | 0.40 | 0.0116 | -0.32 | 0.11 | 0.2385 | -0.11 | 0.01 | 0.6949 |
| N1 | -0.77 | 0.59 | 0.0008 | -0.75 | 0.56 | 0.0013 | -0.72 | 0.52 | 0.0024 | -0.31 | 0.09 | 0.2657 | -0.34 | 0.12 | 0.2130 |
| N2 | **-0.84** | **0.70** | **0.0001** | **-0.84** | **0.70** | **0.0001** | -0.77 | 0.59 | 0.0009 | -0.52 | 0.27 | 0.0452 | -0.21 | 0.04 | 0.4483 |
| N3 | -0.66 | 0.44 | 0.0073 | -0.64 | 0.41 | 0.0105 | -0.60 | 0.36 | 0.0191 | -0.20 | 0.04 | 0.4699 | -0.09 | 0.01 | 0.7617 |
| N4 | -0.55 | 0.30 | 0.0337 | -0.50 | 0.25 | 0.0589 | -0.40 | 0.16 | 0.1412 | -0.34 | 0.12 | 0.2105 | 0.41 | 0.17 | 0.1313 |
| N5 | -0.51 | 0.26 | 0.0529 | -0.56 | 0.31 | 0.0314 | -0.52 | 0.27 | 0.0459 | -0.22 | 0.05 | 0.4249 | -0.32 | 0.10 | 0.2406 |
| N6 | -0.63 | 0.40 | 0.0113 | -0.63 | 0.40 | 0.0111 | -0.55 | 0.30 | 0.0343 | -0.22 | 0.05 | 0.4407 | -0.15 | 0.02 | 0.5836 |
| N7 | 0.62 | 0.39 | 0.0134 | 0.63 | 0.39 | 0.0125 | 0.59 | 0.35 | 0.0207 | 0.29 | 0.08 | 0.2975 | 0.28 | 0.08 | 0.3089 |
| **Negative Image Reports** | E | | | LCC | | | LSC | | | LCCz | | | LSCz | | |
| PANSS Negative Subscale | R | $R^2$ | p | R | $R^2$ | p | R | $R^2$ | p | R | $R^2$ | p | R | $R^2$ | p |
| Total | -0.64 | 0.41 | 0.0019 | **-0.75** | **0.56** | **0.0001** | **-0.69** | **0.47** | **0.0006** | **-0.72** | **0.51** | **0.0003** | **-0.82** | **0.67** | **0.0000** |
| N1 | -0.66 | 0.43 | 0.0011 | **-0.73** | **0.53** | **0.0002** | **-0.71** | **0.50** | **0.0004** | -0.63 | 0.40 | 0.0022 | **-0.73** | **0.54** | **0.0002** |
| N2 | -0.68 | 0.46 | 0.0008 | **-0.74** | **0.55** | **0.0001** | **-0.70** | **0.49** | **0.0004** | -0.64 | 0.41 | 0.0017 | **-0.69** | **0.47** | **0.0006** |
| N3 | -0.65 | 0.42 | 0.0014 | **-0.72** | **0.52** | **0.0002** | **-0.72** | **0.52** | **0.0002** | -0.65 | 0.42 | 0.0014 | **-0.75** | **0.56** | **0.0001** |
| N4 | -0.58 | 0.33 | 0.0061 | **-0.69** | **0.48** | **0.0005** | -0.55 | 0.30 | 0.0097 | -0.61 | 0.38 | 0.0030 | -0.63 | 0.39 | 0.0024 |
| N5 | -0.36 | 0.13 | 0.1073 | -0.48 | 0.23 | 0.0269 | -0.40 | 0.16 | 0.0715 | -0.55 | 0.30 | 0.0095 | -0.66 | 0.44 | 0.0011 |
| N6 | -0.68 | 0.46 | 0.0007 | **-0.76** | **0.57** | **0.0001** | **-0.74** | **0.55** | **0.0001** | -0.63 | 0.39 | 0.0024 | **-0.80** | **0.64** | **0.0000** |
| N7 | 0.49 | 0.24 | 0.0237 | 0.43 | 0.18 | 0.0528 | 0.45 | 0.20 | 0.0399 | 0.07 | 0.00 | 0.7723 | 0.13 | 0.02 | 0.5727 |

**Table 4:** Statistical comparison of connectedness attributes (E, LCC, LSC, LCCz, LSCz) between diagnostic groups (Schizophrenia = S, Bipolar = B, Control = C). Kruskal-Wallis test (SxBxC, Bonferroni corrected for 2 comparisons, p < 0.0250 in bold); Wilcoxon-Ranksum test (SxB, SxC, SxB+C, BxC; Bonferroni corrected for 6 comparisons, p < 0.0083 in bold).

| Ranksum | Comparison | E | LCC | LSC | LCCz | LSCz |
|---|---|---|---|---|---|---|
| **Dream** | SxB | **0.0056** | **0.0031** | **0.0040** | 0.1893 | 0.1893 |
| | SxC | **0.0042** | **0.0045** | **0.0079** | 0.2652 | 0.1239 |
| | Sx(B+C) | **0.0021** | **0.0019** | **0.0031** | 0.1872 | 0.1013 |
| | BxC | 0.5418 | 0.8640 | 0.8642 | 0.9029 | 0.5419 |
| **Negative** | SxB | **0.0081** | 0.0181 | 0.0205 | 0.3418 | 0.1300 |
| | SxC | **0.0009** | **0.0022** | **0.0009** | 0.1421 | **0.0033** |
| | Sx(B+C) | **0.0005** | **0.0015** | **0.0008** | 0.1446 | **0.0060** |
| | BxC | 0.7997 | 0.6719 | 0.8823 | 0.7513 | 0.4856 |
| **Kruskal-Wallis** | Comparison | E | LCC | LSC | LCCz | LSCz |
| **Dream** | S x B x C | **0.0070** | **0.0074** | **0.0112** | 0.3976 | 0.2240 |
| **Negative** | S x B x C | **0.0021** | **0.0056** | **0.0034** | 0.3197 | **0.0158** |

**Table 5:** Statistical comparison of Fragmentation Index between diagnostic groups (Schizophrenia = S, Bipolar = B, Control = C), considering dream + negative image reports, negative image reports or dream reports and applying the Fragmentation Index from dream reports to an independent cohort of chronic psychotic sample [9]. Kruskal-Wallis test (SxBxC, Bonferroni corrected for 3 comparisons p < 0.0167 in bold); Wilcoxon-Ranksum test (SxB, SxC, Sx(B+C), BxC; Bonferroni corrected for 3 comparisons, p < 0.0083 in bold).

| Ranksum | Comparison | Fragmentation Index |
|---|---|---|
| Dream + Negative | SxB | **0.0003** |
| | SxC | **0.0030** |
| | Sx(B+C) | **0.0008** |
| | BxC | 0.4792 |
| Negative | SxB | 0.0221 |
| | SxC | **0.0013** |
| | Sx(B+C) | **0.0011** |
| | BxC | 0.7513 |
| Dream | SxB | 0.0205 |
| | SxC | 0.0137 |
| | SxB+C | **0.0073** |
| | BxC | 0.6330 |
| Dream - Chronic Sample | SxB | **0.0077** |
| | SxC | **2.92E-05** |
| | SxB+C | **0.0001** |
| | BxC | **0.0060** |
| Kruskal-Wallis | Comparison | Fragmentation Index |
| Dream | S x B x C | **0.0025** |
| Negative | S x B x C | **0.0044** |
| Dream | S x B x C | **0.0140** |
| Dream - Chronic Sample | S x B x C | **2.25E-05** |

**Table 6:** Classification quality to sort Schizophrenia patients from others subjects or to sort between low and high negative symptom severity using only the Fragmentation Index of dream + negative image reports, the Fragmentation Index from negative image reports, or the Fragmentation Index from dream reports. Also shown is the independent validation using Fragmentation Index from dream reports applied to a chronic psychotic sample [9].

| Fragmentation Index | Classification | Sensitivity | Specificity | Precision | Recall | F-Measure | AUC | Accuracy |
|---|---|---|---|---|---|---|---|---|
| Dream + Negative | Sx(B+C) | 0.89 | 0.35 | 0.88 | 0.89 | 0.88 | 0.89 | 88.89 |
| | High x Low | 0.93 | 0.13 | 0.94 | 0.93 | 0.93 | 1.00 | 93.33 |
| Only Negative | Sx(B+C) | 0.79 | 0.43 | 0.77 | 0.79 | 0.77 | 0.78 | 78.57 |
| | High x Low | 0.91 | 0.10 | 0.91 | 0.91 | 0.91 | 0.97 | 90.48 |
| Only Dream | Sx(B+C) | 0.78 | 0.38 | 0.80 | 0.78 | 0.79 | 0.81 | 77.78 |
| | High x Low | 0.80 | 0.20 | 0.82 | 0.80 | 0.80 | 0.78 | 80.00 |
| Dream - Chronic Sample | Sx(B+C) | 0.82 | 0.32 | 0.82 | 0.82 | 0.81 | 0.80 | 81.67 |
| | High x Low | 0.75 | 0.38 | 0.75 | 0.75 | 0.74 | 0.84 | 75.00 |

**Figures**

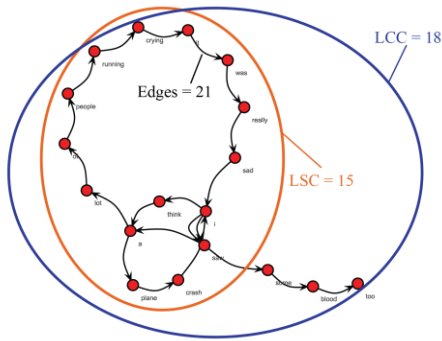
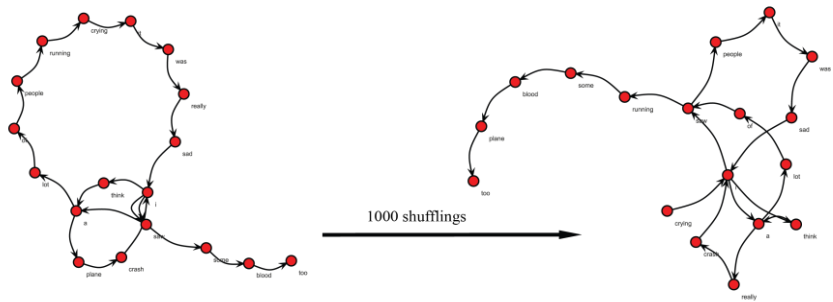
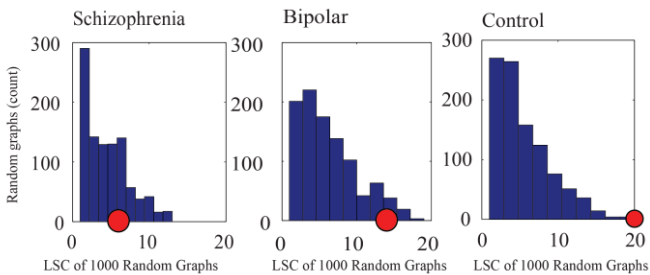
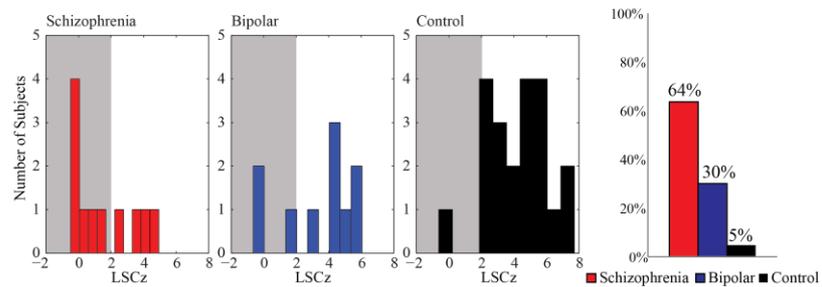

**Figure 1: Speech Graph Connectedness attributes and Random-like connectedness in Schizophrenia.** A) Illustrative example of a text represented as a graph, showing connectedness attributes Edges, LCC and LSC. B) Illustrative example of random graphs created from an original report. By shuffling the word order, the 1000 random graphs maintained the same words, but with a random structure. C) Examples of one negative image report compared to 1,000 random graphs for each group. Estimation of original LSC (red dot) distance from a 1,000 random graph distribution (blue histogram) by zscore - LSCz. D) LSCz histogram from each diagnostic group, considering as random-like speech those with LSCz = -2 until 2 (2 standard deviation from a random graph distribution); and the percentage of random-like speech in each diagnostic group.

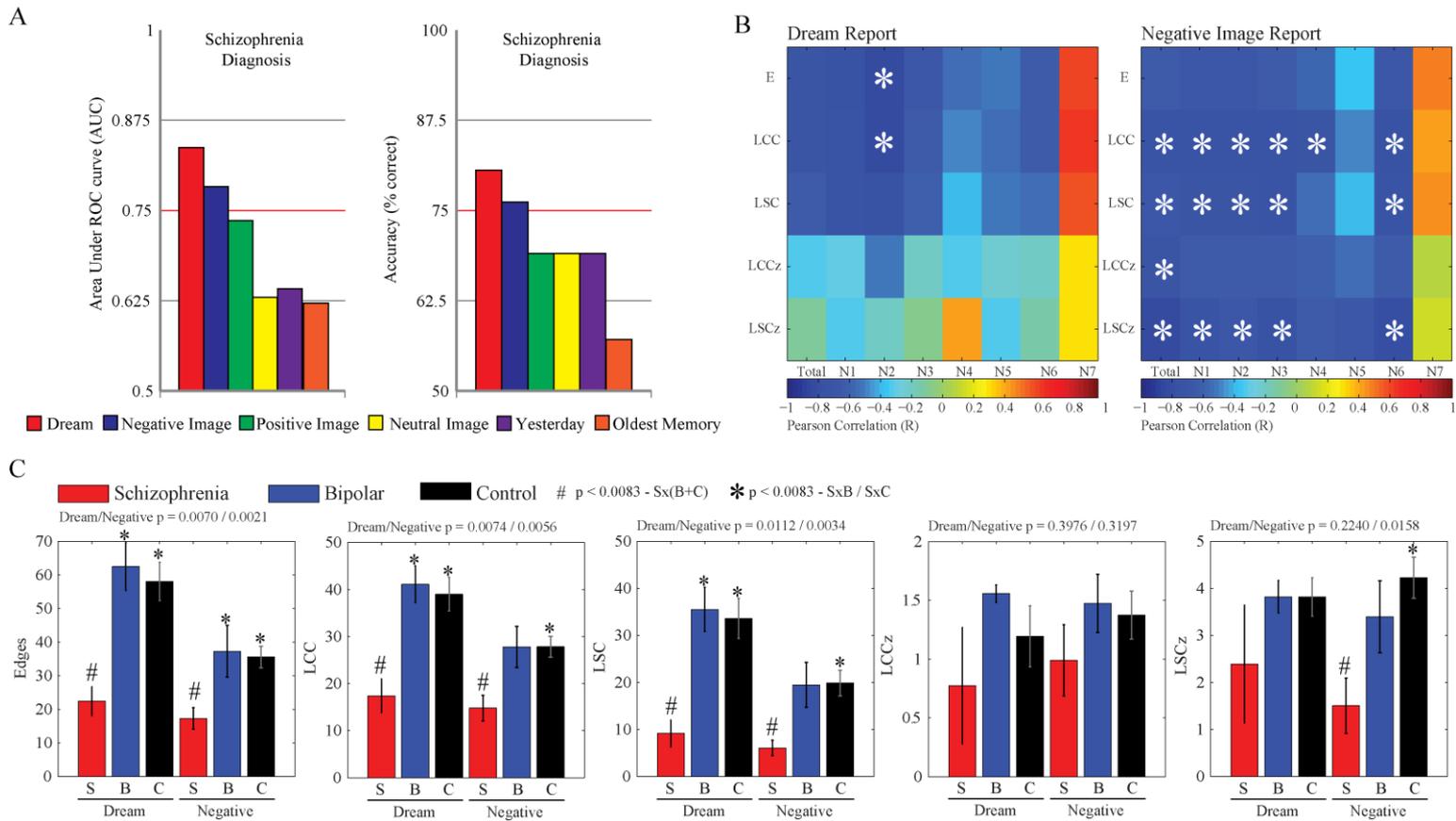

**Figure 2: While reporting a dream or a negative image, a less connected and random-like speech is more frequently observed in Schizophrenia group and it is correlated with negative symptoms measured by PANSS.** A) Schizophrenia diagnostic classification using 5 connectedness attributes (E, LCC, LSC, LCCz and LSCz) using 6 time-limited memory reports. Only dream and negative image reports classified Schizophrenia group versus Bipolar and Control group with AUC > 0.75 and accuracy > 75%. B) Pearson correlation matrix between connectedness attributes and PANSS negative subscale for dream and negative image reports (white * means p < 0.0006 and color bar indicates R values). C) Connectedness attributes from dream and negative image reports compared between groups (Kruskal-Wallis tests: p value for dream / negative image reports indicated in each title; Wilcoxon-Ranksum test: # means p < 0.0083 – Schizophrenia versus Bipolar and Control groups, * means p < 0.0083 – Schizophrenia versus Bipolar or Control groups).

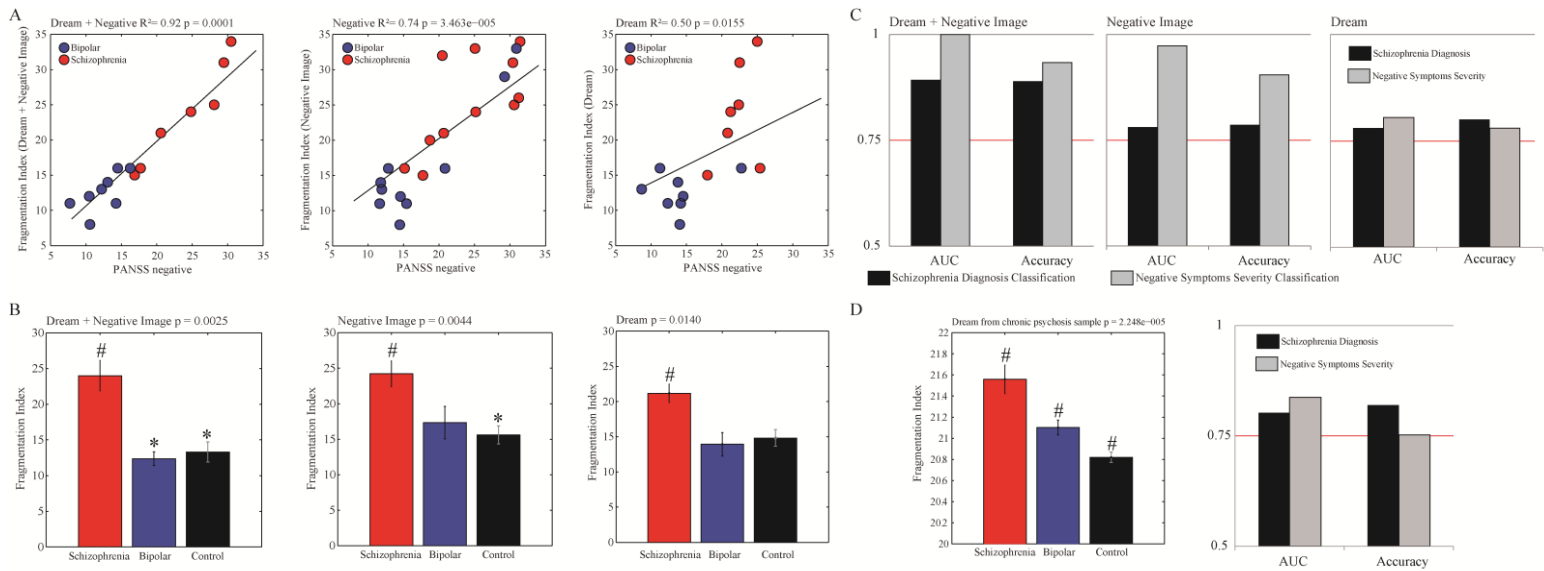

**Figure 3: Fragmentation Index classifies negative symptoms severity and predicts Schizophrenia diagnosis 6 months in advance.** A) Multilinear correlation between PANSS negative subscale versus Fragmentation Index from dream + negative image reports, from negative image reports, or from dream reports ($R^2$ and p value indicated on title; linear coefficients used to calculate Fragmentation Index on Findings). B) Bar plot of the mean and standard error of Fragmentaion Index from dream + negative image reports, from negative image reports, or from dream reports for diagnostic groups (schizophrenia in red, bipolar in blue and control in black; Kruskal-Wallis tests: p value indicated in each title; # indicates p < 0.0083 - Schizophrenia > Bipolar and Control groups; * indicates p < 0.0083 - Schizophrenia > Bipolar or Control groups). C) Classification quality using only Fragmentation Index from dream + negative image reports, from negative image reports, or from dream reports (measured by area under ROC curve (AUC) and Accuracy –prediction of Schizophrenia diagnosis 6 months in advance (in black); Negative Symptom Severity measured by PANSS negative subscale (in gray). D) Validation of Fragmentation Index from dream reports only distinguishes groups (Schizophrenia, Bipolar and Control) when applied to an independent cohort of chronic psychotic patients [9], using statistical comparison (Schizophrenia in red, bipolar in blue and control in black; Kruskal-Wallis tests: p value indicated in each title; # indicates p < 0.0083 - Schizophrenia > Bipolar and Control groups; * indicates p < 0.0083 - Schizophrenia > Bipolar or Control groups) and classification quality (measured by AUC and Accuracy – prediction of Schizophrenia diagnosis 6 months in advance (in black); Negative Symptom Severity measured by PANSS negative subscale (in gray)).

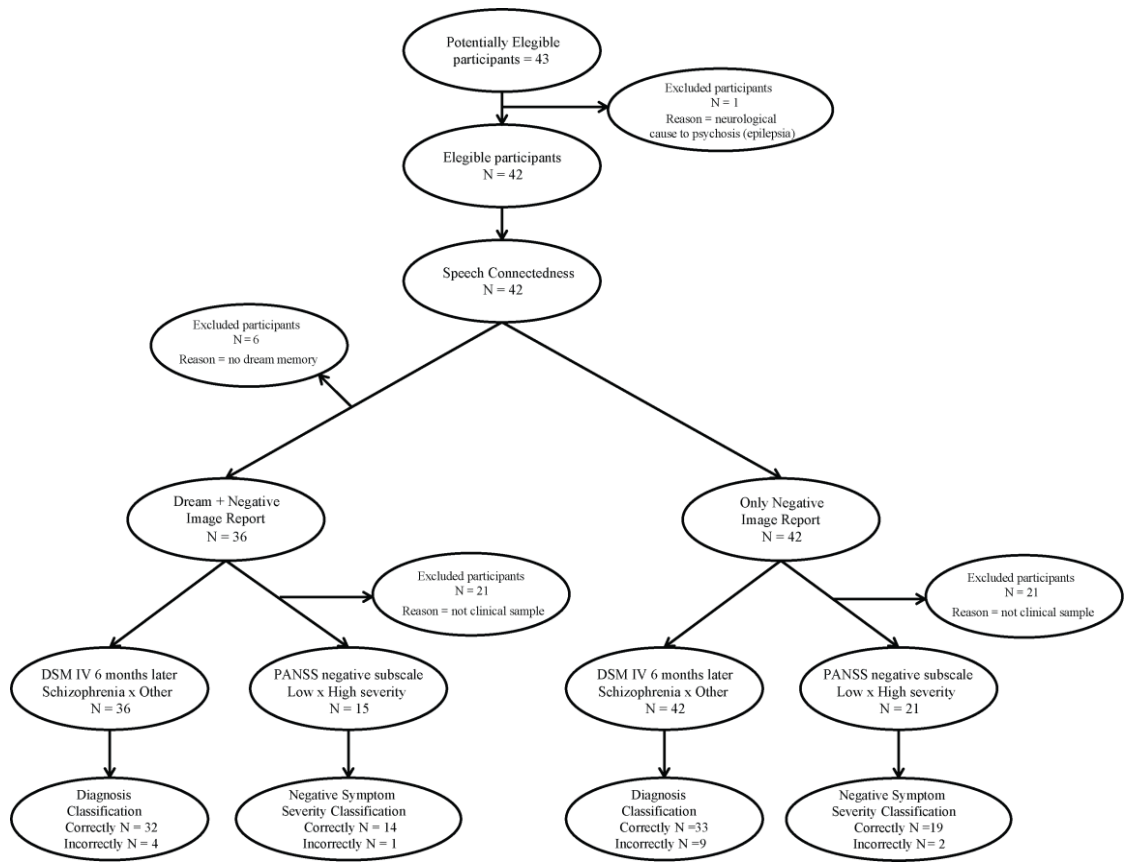

**Figure 4: Illustrative diagram of the flow of participants through the study**

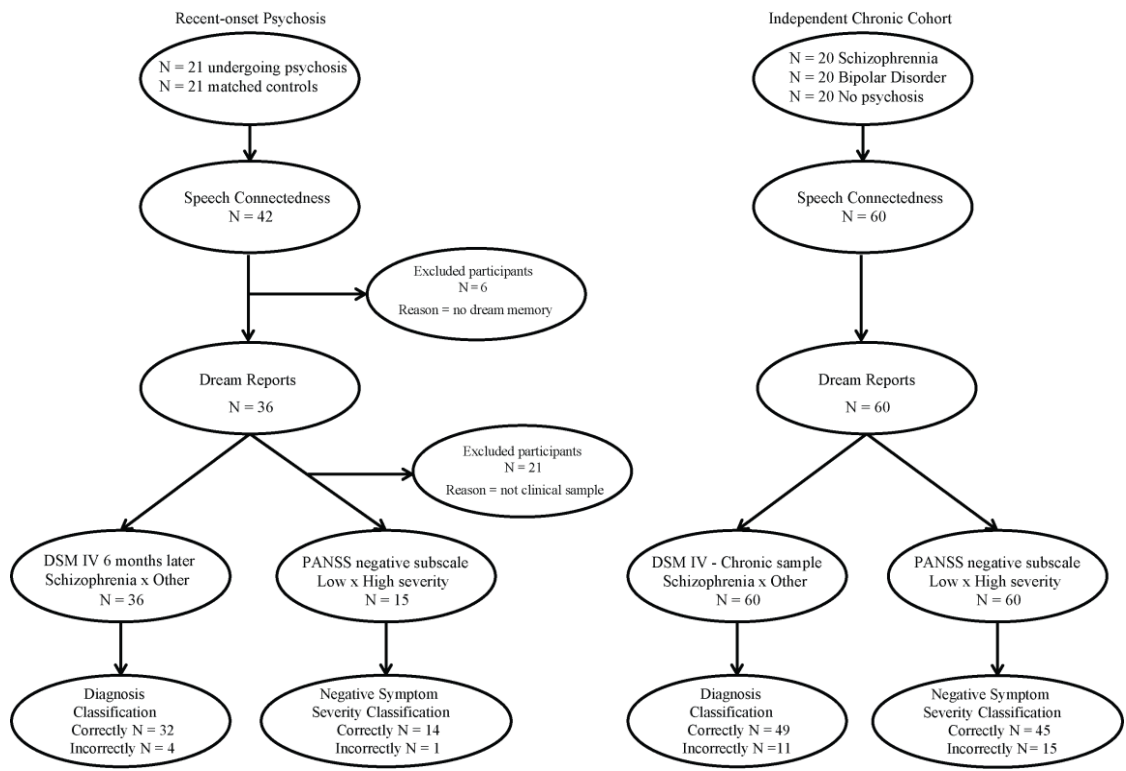

**Figure 5: Illustrative diagram of the flow of participants through validation in an independent cohort chronic psychosis sample.**